# Resonant optical gating of suspended carbon nanotube transistor

*R. McCoy,[1] F. Anderson,[2] E. L. Carter,[1] R. L. Smith[2]*

[1]Sullivan University, Louisville, KY 40205, USA

[2]Rio Salado College, Tempe, AZ 85281, USA

**Abstract:** *Building smaller transistors with enhanced functionality is critical in extending the limits of Moore's law and meeting the demands of the electronics industry. Here we demonstrate transistor operation in a suspended single carbon nanotube (CNT) using feedback-enabled radiation pressure of a near-field focused laser that enabled significant changes in conductivity of the CNT. Further, using in-situ tip-enhanced Raman spectroscopy, we show that the change in conductivity of over five orders in magnitude is accompanied by self-induced defect states within the CNT. The entire structure is less than 10 nm in dimension and shows promise of scalability. This is a novel method for achieving logic operations at the nanoscale.*

The electronics industry is turning towards nanotechnology to meet the growing need to extend Moore's law into the regime of operation and scalability that cannot be achieved by conventional semiconductor technology. Apart from the physical scale of electronic gadgets, there is a need to enhance their functionality to pack more features within a smaller space. This puts great emphasis on nanoscale materials that can provide such dense functionality. Transistors enable current logic operation and also several other electronic operations including amplification, memory, etc. Hence, making smaller and better transistors is of prime importance and focus of the nanotechnology community. Carbon nanotubes (CNT) are arguably the most widely used nanomaterial in a range of applications ranging from electronics, military, reinforced structures, optics, etc. The ease of fabrication of CNT coupled with the immense research that has been devoted to its application to several domains makes it easier to extend its utility to electronics. Here we demonstrate the utility of CNTs in a single-CNT transistor. In a novel approach to gate the transistor (control its conductivity), we utilized a near-field focused optical laser interferometer-like setup that produced significant radiation pressure, sufficient to physically bend the CNT, thereby altering its conductivity. We tuned the gating process, using feedback, to exploit the resonant frequency response of the CNT, which produced a large signal sufficient to obtain real-time feedback. We showed that the conductivity changed by over five orders of magnitude as the gating process was controlled (conveniently read by an effective gating voltage). Further, using tip enhanced Raman spectroscopy (TERS), we showed that the defect states in the CNT appeared to increase and decrease with altering of the conductivity. This process provides an insight into the mechanism that achieves such a large change in conductivity. Also, this helps us speculate on the required conditions that enable such transistor behavior in structures that are driven by several other stimuli, including conventional gating through a dielectric.

The growth of CNT was performed by controlled annealing of metallic catalyst precursors. A lightly boron-doped Si wafer was chosen as the substrate with 500 µm thickness. This was cleaned using a standard RCA cleaning process, in order to remove contaminants, ionic impurities, residual oxides, etc. To drive moisture out, the substrates were baked in vacuum at ~450 K for two hours. Following this, we deposited 5 nm of Al and 5 nm of Fe using e-beam evaporation. We then cleaned the substrates using acetone and iso-propyl alcohol, along with dry nitrogen gas. The substrate was the taken through a catalytic growth process, wherein we introduced the following gases: Ar gas (an inert gas) at room temperature (25 C) for 30 minutes, using Ar partial pressure of 1000 mTorr; ramping up of the temperature to 760 C over 15 minutes without changing the gas environment; flow of ethylene gas at a partial pressure of 450 mTorr along with hydrogen gas at a partial pressure of 150 mTorr at 760 C; Ar flow at a partial pressure of 1000 mTorr at 760 C for 10 minutes; followed by cool down. The details of this process is described elsewhere in our prior publications. In order to bridge the CNT as a cantilever between two electrodes with a gap in between, the metallic catalysts were patterned using electron beam lithography on either of the electrodes, after which the CNT was allowed to grow across the gap in the process described above. A sufficiently small feature of the catalyst allowed for the growth of a single CNT. This naturally resulted in a cantilever that was anchored either at one end or both ends.  Figure 1a is a schematic of the setup described above. This also depicts the scheme used for optical gating using radiation pressure. Figure 1b is a scanning electron micrograph (SEM) of CNTs suspended on one end and anchored on the other. Figure 1c is SEM of a single CNT anchored on both ends and suspended in between the anchoring surfaces.

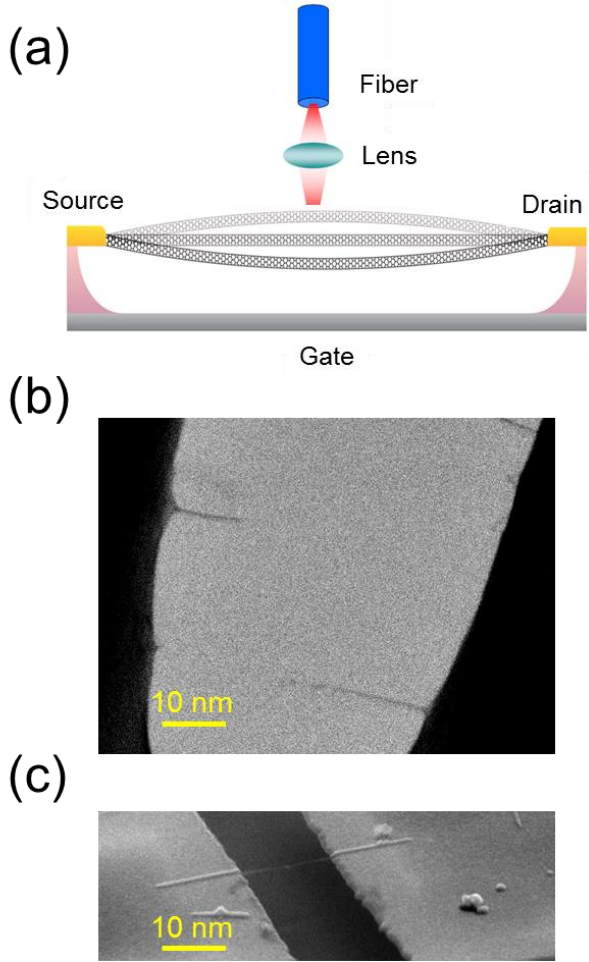

Figure 1: (a) Schematic of the suspended CNT transistor and its gating process using radiation pressure. (b) Scanning electron micrograph (SEM) of CNTs suspended and anchored on one end. (c) SEM of a single CNT suspended across a gap between two platinum electrodes and anchored on two ends.

The mechanism of actuation, or gating, is depicted in Figure 2a. Here we used a reference photodiode which produced a laser source, coupled with a readout laser at 1310 nm through a fiber coupler. There was a signal photodiode which went through an amplifier followed by a phase shifter which controlled a 1550 nm laser. The 1310 nm laser and 1550 nm laser were combined by a wavelength division multiplexer (WDM) which was then focused through a lens, followed by a near-field tip (not pictured), which was then focused onto the CNT cantilever. The feedback part of this system corresponds to exciting or damping the frictional component of the damped harmonic oscillator. One such system can be described by the equation

$$x(\omega) = \frac{\omega_o^2/k}{\omega_o^2 - \omega^2 + i\Gamma\omega}[F_{\text{thermal}}(\omega) + F_{\text{ext}}(\omega)],$$
$$\dots (1)$$

Where ω is the oscillation frequency and F indicates force on the cantilever. Γ is the feedback component affecting the first derivative of displacement, x. Such a system is capable of constantly tuning the resonance of the cantilever, especially its amplitude without affecting its Q factor. The Q factor in our case was as high as 100,000. To prevent the highly oscillatory nature of the change in conductivity, we constantly damped the cantilever. Damping of the cantilever without reducing its Q factor enabled us to achieve high signal sensitivity without compromises. It is notable that such a system was primarily invented to study effects of gravity on cantilevers.

Figure 2b depicts the signal read-out mechanism, where we used multiple comparators that allowed us to obtain a reference signal from the background (laser), which was sued as the source of background noise. This was then subtracted from the signal obtained from the read-out laser. This provided a very high signal level. Figure 2c displays data from the resonance of the cantilever before damping of the same. This shows a very clearly defined resonant frequency in the order of several hundred kHz. The resolution from this read-out technique was found to be 0.01 nm. We fitted a sum of two lorentzian peaks to the frequency response, descried elsewhere. From this system, we define the voltage produced by the laser read-out as the gate voltage on the CNT.

In order to simulate the observed transistor behavior, we employ traditional semiconductor transistor equations. For a long channel field effect transistor, without substrate bias, the drain current can be expressed as:

$$I_D = \mu_{eff} C_{ox} \frac{W_{eff}}{L_{eff}} [(V_G - V_{TH})V_D - \frac{1}{2}V_D^2] \dots (2)$$

Where $\mu_{eff}$, $W_{eff}$ and $L_{eff}$ are the effective channel mobility, effective channel length, and effective channel width which will be discussed later. $C_{ox}$ is taken to be $3.9 \times 10^{-7}$ F/cm$^2$, which is taken from the direct measurements. When $V_D$ starts from a small value, Eq (1) can be reduced to $I_D = \mu_{eff} C_{ox} \frac{W_{eff}}{L_{eff}} [(V_G - V_{TH})V_D]$ the $I_D$ goes up with $V_D$ linearly and this corresponds to the triode region of $I_D$-$V_D$ curves.

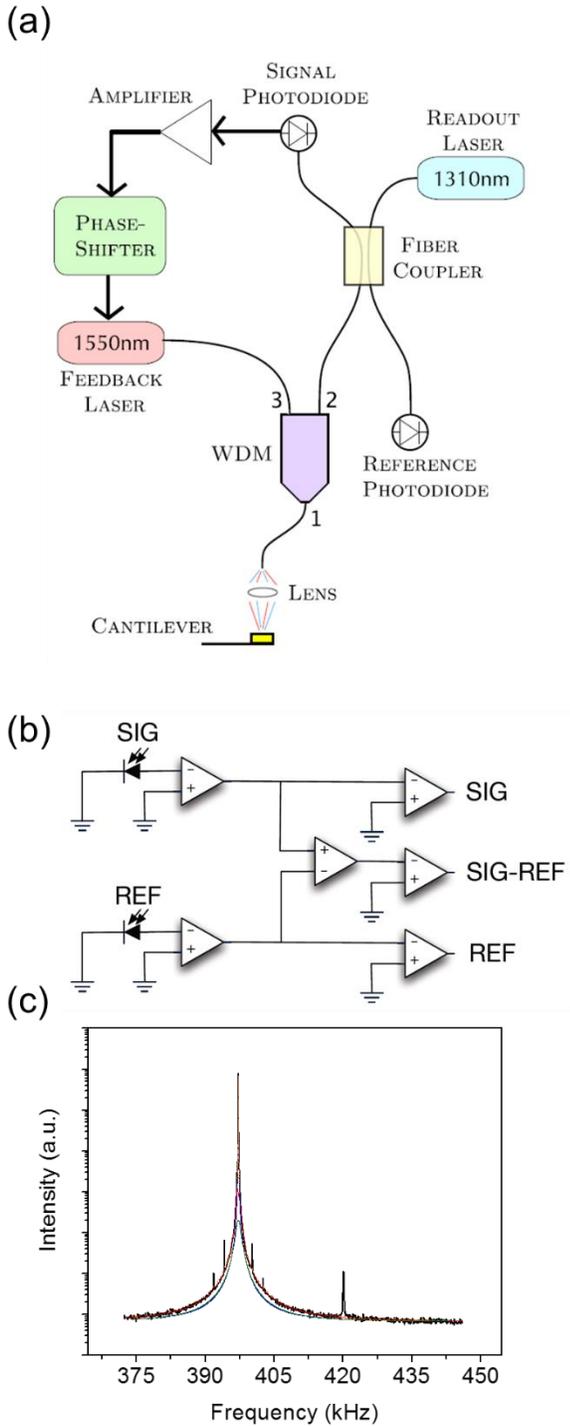

Figure 2: (a) Schematic of the gating setup. This shows the laser focus setup, feedback setup, wavelength division multiplexing and the CNT cantilever. (b) Signal readout, by incorporating a reference (background/noise) subtraction. (c) Resonance response of one of the cantilevers (suspended CNT) under study.

As $V_D$ increases furthermore, the $\frac{1}{2}V_D^2$ term plays a more important role and finally when $V_D$ reaches the pinch-off point, it will be saturated at $V_G$-$V_{TH}$, i.e. $V_{D,sat}$= $V_G$-$V_{TH}$. The saturation drain current thus becomes $I_{D,sat} = \mu_{eff}C_{ox}\frac{W_{eff}}{L_{eff}}\left[\frac{1}{2}(V_G - V_{TH})^2\right]$ and it refers to the saturation region where $I_D$ is in independent of $V_D$. Increasing $V_G$ intuitively makes $I_D$ larger, as well as the value at which $V_D$ reaches the saturation point. Now let's look at the $I_D$-$V_G$ curves. Before $V_G$ reaches $V_{TH}$, the inversion layer hasn't formed so there is barely any $I_D$. As $V_G > V_{TH}$, the channel is turned on and $I_D$ follows the above equation, goes up with larger $V_G$. Increasing $V_D$, again, intuitively makes $I_D$ larger; however, when $V_D$ is larger than $V_G$-$V_{TH}$, saturation happens and the curve will no longer be moved upward.

The threshold voltage, $V_{TH}$, can be obtained with equations above in the different regimes. Plotting $I_D$-$V_G$ curves with small $V_D$, which ensures the curves are located in the linear region, we can get $V_{TH}$ as the x-intercept minus 1/2 $V_D$ of the line extrapolated by the linear part of the curves. To double check the values, vertical lines are drawn in Figure 3a from $V_{TH}$ to confirm that they correspond to the linear part of $I_D$-$V_D$ curves.

In comparison to the $V_{TH}$ from simulation results, we can see that for the cross section, the values we got are both around -0.75 V. However, the $V_{TH}$ for our devices were simulated to be negative rather than positive. It can be attributed to the existence of an effective "P-well". The formation of P-well adds more uncertainty throughout the process via parameters such as doping concentration, oxide thickness, interface charge, etc. They are all possible sources of inaccuracy and can be added up to the deviation of simulation model itself. On the other hand, we use the CNT as channel, so it is understandable that the $V_{TH}$ from simulation is closer to the empirical values.

A method to extract effective mobility, $\mu_{eff}$, using $I_D$-$V_G$ curves is proposed by Gupta et.al. In this method, $\mu_{eff}$ in the linear region is given by $\mu_{eff} = \frac{g_m L_{eff}}{W_{eff}C_{ox}V_D}$ where $g_m$ is the transconductance which will be introduced in the next section. In the saturation region, $\mu_{eff}$ is given by $\mu_{eff} = \frac{2L_{eff}}{W_{eff}C_{ox}}\left(\frac{\partial\sqrt{I_D}}{\partial V_G}\right)^2$. The $\mu_{eff}$ obtained by this method is plotted in Figure 3c, which shows a considerable inconsistency with the first method we used.

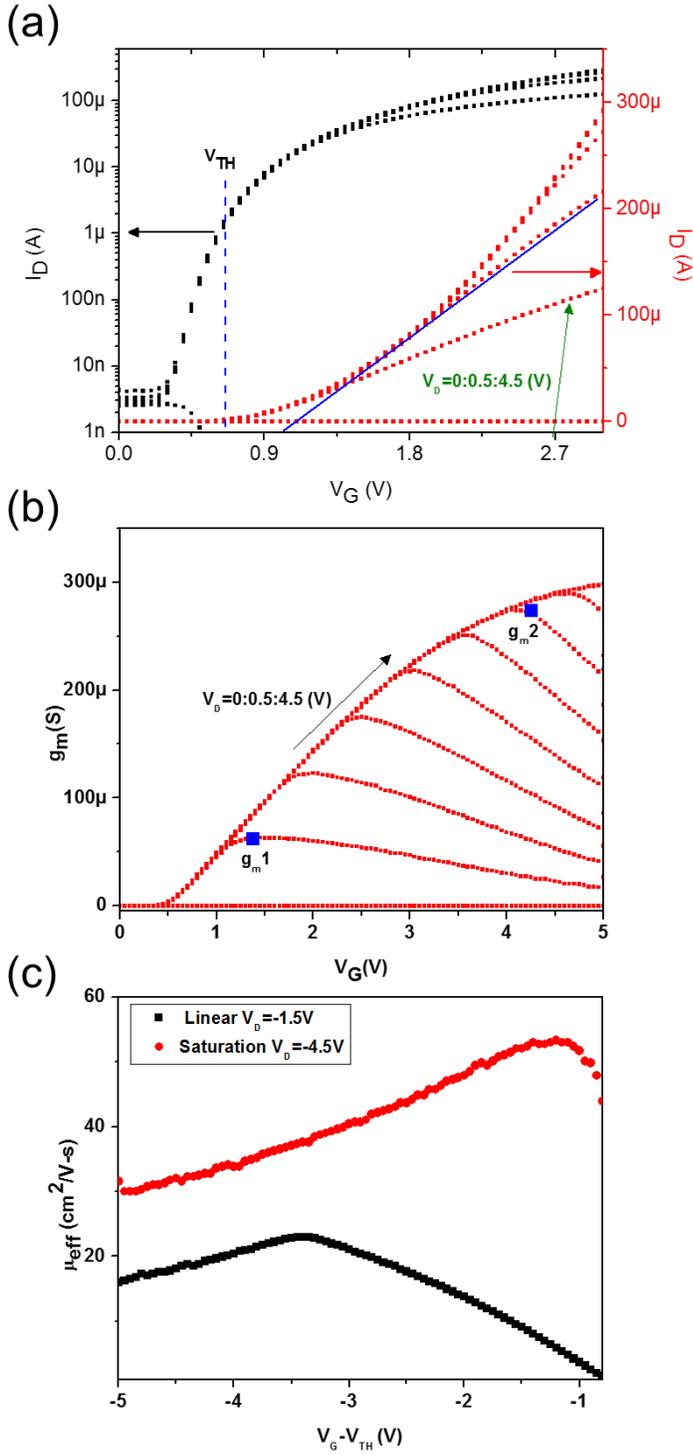

Figure 3: (a) Transistor characteristics by plotting the gate voltage (defined in the text) versus the drain current. The drain current is displayed in both linear and logarithmic scales (on the left and right ordinates). The applied drain

voltage is indicated. (b) Transconductance as a function of gate voltage and drain voltage. (c) Effective mobility calculated for two different drain voltages as a function of gate voltage.

The reason for this inconsistency between traditional transistor modeling and CNT transistor modeling can be found to be well explained in elsewhere. The effective mobility observed in this method follows the equation

$$\mu^* = \mu_n \left( 1 - \frac{1}{\sqrt{1 + \frac{2(2\phi_F + V_{SB})C_{ox}^2}{qN\varepsilon_s}}} \right) \ \ldots (3)$$

Where $\mu^*$ is the modified mobility in the variable depletion layer charge model. Quadratic model yields a larger drain current compared to the more accurate variable depletion layer charge model as described in the reference.

We can represent a biased CNT transistor by an equivalent circuit where changes in drain current are related to gate voltage changes by $g_m = \frac{\partial I_D}{\partial V_G}$ . By taking the slopes of $I_D$-$V_G$ curves, as illustrated in Figure 3, $g_m$ are extracted and plotted in Figure 3b. The transconductance is zero when $V_G$ is below $V_{TH}$ because there is little drain current. It goes through a maximum at the point of inflection of the $I_D$-$V_G$ curve, and then decreases. This decrease is due to two factors. The first one is degradation of the effective channel mobility as a function of increasing transverse electric field across the gate oxide. The mobility of carriers in the channel of a CNT transistor is lower than in bulk semiconductors because there are additional scattering mechanisms such as surface roughness. This mobility degradation increases with the gate bias because a higher gate bias draws the carriers closer to the oxide-silicon interface, where they are more influenced by the interfacial roughness. The second one is source/drain series resistance. For a certain applied drain bias to the source/drain terminals, part of the applied voltage is wasted as an ohmic voltage drop across these resistances, depending on the drain current (or gate bias).In practice, $g_m$ parameter plays a role as the device output admittance or the a.c. conductance of the channel between the source and drain. Higher $g_m$ is mostly pursued. As channel length decreases, the $g_m$ goes up and the other way around for the channel width reduction. These trends are consistent with equations presented above. We extract the maximum $g_m$ and also the gate voltage at which this maximum occurs since we feel it's important for the device to be labeled for these numbers.

In order to understand the nanoscale chemical changes to the CNT during transistor operation, we performed in-situ tip enhanced Raman spectroscopy (TERS), which has a spatial resolution of ~1 nm to obtain the Raman spectrum of the CNT. The spectra corresponding to the two prominent states of the transistor, namely, the low resistance state (high gate voltage) and high resistance state (no gate voltage) are presented in Figure 4. The spectra display the known modes of vibration, stretching and rotations of the CNT. The most prominent difference between the two spectra is the growth of the defect peak (labeled 'D') in the OFF state of the transistor. This shines light on the local changes to the structure of the CNT that enables the massive change in resistance. The reduction in the intensity of the defect peak indicates relaxation of the bonds and promotes higher conductivity by lesser scattering of the electronic conduction by the defect sites. This also helps us understand the location of the conduction sites within the CNT and identifies them as predominantly being situated on the surface of the CNT walls.

In conclusion, we constructed a transistor stucture using a single suspended CNT and enabled transistor behavior by employing a sophisticated but scalable resonantly tuned optical gating technique. We show changes in conductivity by several orders of magnitude upon application of a gate signal (measured in effective gate voltage). We further shine light on the local chemistry in action during transistor operation by utilizing in-situ TERS measurements.

The authors gratefully acknowledge Prof. K. Ashby for fruitful discussions and comments. The authors also acknoledge the fabrication facility at the Arizona State University for fabrication of the CNT transistor structures.

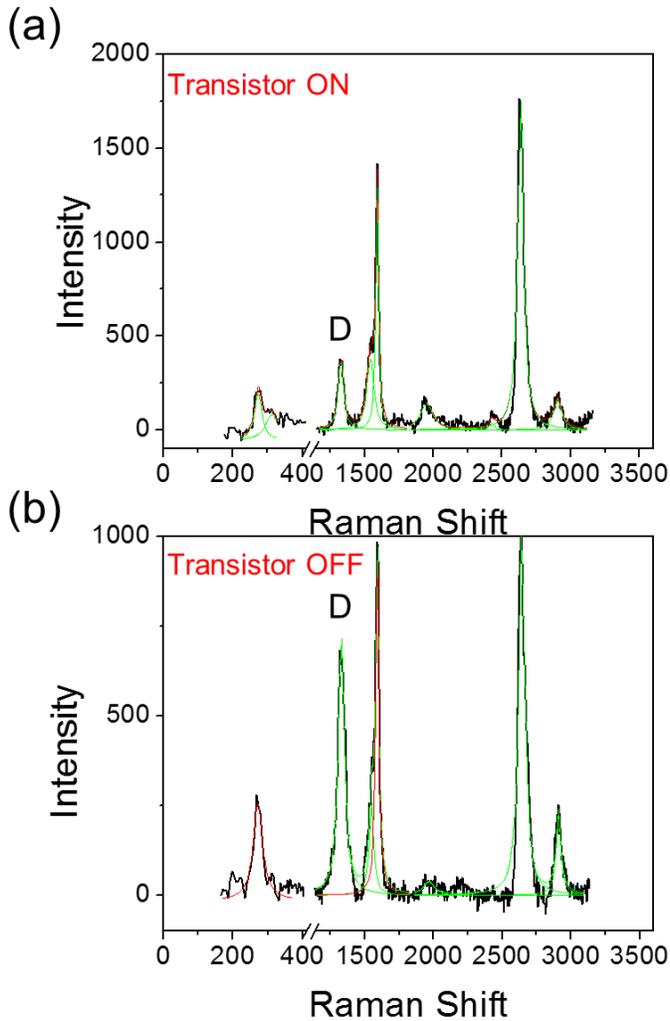

Figure 4: Tip enhanced Raman spectroscopy of the CNT in the two states of the transistor, namely, ON (low resistance) and OFF (high resistance). The peak labeled 'D' is the defect peak that shows significant changes upon switching of the conductance of the CNT.

### References

[1]  Maultzsch, J., S. Reich, C. Thomsen, H. Requardt, and P. Ordejón, 2004, "Phonon Dispersion in Graphite," Phys. Rev. Lett. **92**, 075501.

[2]  McCann, E., 2006, "Asymmetry gap in the electronic band structure of bilayer graphene," Phys. Rev. B **74**, 161403.

[3]  McCann, E., and V. Fal'ko, 2006, "Landau-Level Degeneracy and Quantum Hall Effect in a Graphite Bilayer," Phys. Rev. Lett. **96**, 086805.

[4]  MacDonald, A. H., 1990, *Quantum Hall Effect: A Perspective* (Kluwer Academic, Dordrecht, The Netherlands).

[5]  Mak, K. F., C. H. Lui, J. Shan, and T. F. Heinz, 2009, "Observation of an Electric-Field-Induced Band Gap in Bilayer Graphene by Infrared Spectroscopy," Phys. Rev. Lett. **102**, 256405.

[6]  Mak, K. F., M. Y. Sfeir, Y. Wu, C. H. Lui, J. A. Misewich, and T. F. Heinz, 2008, "Measurement of the Optical Conductivity of Graphene," Phys. Rev. Lett. **101**, 196405.

[7] Dikin, Dmitriy A., Sasha Stankovich, Eric J. Zimney, Richard D. Piner, Geoffrey H. B. Dommett, Guennadi Evmenenko, SonBinh T. Nguyen, and Rodney S. Ruoff, 2007, "Preparation and characterization of graphene oxide paper," Nature (London) **448**, 457.

[8] Dresselhaus, M. S., A. Jorio, L. G. Cançado, G. Dresselhaus, and R. Saito, 2012, "Raman Spectroscopy: Characterization of Edges, Defects, and the Fermi Energy of Graphene and sp2 Carbons," in *Graphene Nanoelectronics: Metrology, Synthesis, Properties and Applications*, edited by Hasan Reza (Springer, Berlin), Chap. 2, p. 15.

[9] Dresselhaus, Mildred S., Ado Jorio, Mario Hofmann, Gene Dresselhaus, and Riichiro Saito, 2010, "Perspectives on Carbon Nanotubes and Graphene Raman Spectroscopy," Nano Lett. **10**, 751.

[10] Wevers, B. M. H. R., 1984, "A study of spiral galaxies using HI synthesis observations and photographic surface photometry," Ph.D. thesis (Groningen University).

[11] Kumar, S, 2012, "Learning from Solar Cells - Improving Nuclear Batteries,", Stanford University, http://large.stanford.edu/courses/2012/ph241/kumar2/

[12] White, S. D. M., and M. J. Rees, 1978, "Core condensation in heavy halos - A two-stage theory for galaxy formation and clustering," Mon. Not. R. Astron. Soc.**183**, 341 [http://adsabs.harvard.edu/abs/1978MNRAS.183..341W].

[13] Widrow, L. M., and J. Dubinski, 2005, "Equilibrium Disk-Bulge-Halo Models for the Milky Way and Andromeda Galaxies," Astrophys. J. **631**, 838.

[14] Widrow, L. M., S. Gardner, B. Yanny, S. Dodelson, and H. Y. Chen, 2012, "Galactoseismology: Discovery of Vertical Waves in the Galactic Disk," Astrophys. J.**750**, L41.

[15] Zaritsky, D., J. E. Colucci, P. M. Pessev, R. A. Bernstein, and R. Chandar, 2012, "Evidence for Two Distinct Stellar Initial Mass Functions," Astrophys. J. **761**, 93.

[16] Zibetti, S., S. Charlot, and H. W. Rix, 2009, "Resolved stellar mass maps of galaxies - I. Method and implications for global mass estimates," Mon. Not. R. Astron. Soc. **400**, 1181.

[17] Zibetti, S., A. Gallazzi, S. Charlot, D. Pierini, and A. Pasquali, 2013, "Near-infrared spectroscopy of post-starburst galaxies: a limited impact of TP-AGB stars on galaxy spectral energy distributions," Mon. Not. R. Astron. Soc. **428**, 1479.

[18] Agustsson, I., and T. G. Brainerd, 2006, "The Orientation of Satellite Galaxies: Evidence of Elongation in the Direction of the Host," Astrophys. J. **644**, L25.

[19] Aihara, H., C. Allende Prieto, and D. e. An, 2011, "The Eighth Data Release of the Sloan Digital Sky Survey: First Data from SDSS-III," Astrophys. J. Suppl. Ser. **193**, 29.

[20] Alaghband-Zadeh, S., *et al.*, 2012, "Integral field spectroscopy of 2.0<z<2.7 submillimetre galaxies: gas morphologies and kinematics," Mon. Not. R. Astron. Soc.**424**, 2232.

[21] Amorisco, N. C., A. Agnello, and N. W. Evans, 2013, "The core size of the Fornax dwarf spheroidal," Mon. Not. R. Astron. Soc. **429**, L89.

[22] Kumar, S., 2011, "Energy from radioactivity," Stanford University, http://large.stanford.edu/courses/2011/ph240/kumar2/

[23] Ferrari, A. C., *et al.*, 2006, "Raman Spectrum of Graphene and Graphene Layers," Phys. Rev. Lett. **97**, 187401.

[24] Ferrari, Andrea C., 2007, "Raman spectroscopy of graphene and graphite: Disorder, electron-phonon coupling, doping and nonadiabatic effects," Solid State Commun. **143**, 47.

[25] Filleter, T., J. L. McChesney, A. Bostwick, E. Rotenberg, K. V. Emtsev, Th. Seyller, K. Horn, and R. Bennewitz, 2009, Phys. Rev. Lett. **102**, 086102.

[26] Fogler, M. M., F. Guinea, and M. I. Katsnelson, 2008, "Pseudomagnetic Fields and Ballistic Transport in a Suspended Graphene Sheet," Phys. Rev. Lett. **101**, 226804.

[27] Fogler, Michael M., 2004, "Nonlinear screening and percolative transition in a two-dimensional electron liquid," Phys. Rev. B **69**, 121409.

[28] Fogler, Michael M., 2009, "Neutrality Point of Graphene with Coplanar Charged Impurities," Phys. Rev. Lett. **103**, 236801.

[29] Martin, I., and Y. M. Blanter, 2009, "Transport in disordered graphene nanoribbons," Phys. Rev. B **79**, 235132.

[30] Martin, J., N. Akerman, G. Ulbricht, T. Lohmann, J. H. Smet, K. von Klitzing, and A. Yacobi, 2008, "Observation of Electron-hole Puddles in Graphene Using a Scanning Single Electron Transistor," Nature Phys. **4**, 144.

[31] Martin, J., N. Akerman, G. Ulbricht, T. Lohmann, K. von Klitzing, J. H. Smet, and A. Yacoby, 2009, "The nature of localization in graphene under quantum Hall conditions," Nature Phys. **5**, 669.

[32] Kumar, S., 2012, "Fundamental limits to Moore's law," Stanford University, http://large.stanford.edu/courses/2012/ph250/kumar1/

[33] Ebbesen, T. W., 1994, "Carbon nanotubes," Annu. Rev. Mater. Sci. **24**, 235.

[34] Ebbesen, T. W., 1996, "Carbon nanotubes," Phys. Today **49** (6), 26.

[35] Suhas, K., "Modelling Studies and Fabrication Feasibility with LTCC for Non-Conventional RF MEMS Switches," 2007, IISC.

[36] Ebbesen, T. W., and P. M. Ajayan, 1992, "Large-scale synthesis of carbon nanotubes," Nature (London) **358**, 220.

[37] Henrald, L., E. Hernández, P. Bernier, and A. Rubio, 1999, "van der Waals interaction in nanotube bundles: Consequences on vibrational modes," Phys. Rev. B **60**, R8521.

[38] Huffman, D. R., 1991, "Solid C60," Phys. Today **44**, 22.

[39] Iijima, S., 1991, "Helical microtubules of graphitic carbon," Nature (London) **354**, 56.

[40] Krätschmer, W., L. D. Lamb, K. Fostiropoulos, and D. R. Huffman, 1990, "Solid C60: A new form of carbon," Nature (London) **347**, 354.

[41] S. Kumar, A. M. Vadiraj, 2016, "Smart packaging of electronics and integrated MEMS using LTCC", arXiv:1605.01789, http://arxiv.org/abs/1605.01789

[42] Kawamura, T., and S. Das Sarma, 1992, "Phonon-scattering-limited electron mobilities in AlxGa1−xAs/GaAs heterojunctions," Phys. Rev. B **45**, 3612.

[43] Kawamura, T., and S. Das Sarma, 1996, "Spatial correlation effect of ionized impurities on relaxation and scattering times in high-mobility GaAs heterojunctions,"Solid State Commun. **100**, 411.

[44] Kechedzhi, K., O. Kashuba, and V. I. Fal'ko, 2008, "Quantum kinetic equation and universal conductance fluctuations in graphene," Phys. Rev. B **77**, 193403.

[45] Kharitonov, M. Y., and K. B. Efetov, 2008, "Universal conductance fluctuations in graphene," Phys. Rev. B **78**, 033404.

[46] Kim, E., and A. H. Castro Neto, 2008, "Graphene as an electronic membrane," Europhys. Lett. **84**, 57007.

[47] Fang, Zheyu, Sukosin Thongrattanasiri, Andrea Schlather, Zheng Liu, Lulu Ma, Yumin Wang, Pulickel M. Ajayan, Peter Nordlander, Naomi J. Halas, and F. Javier García de Abajo, 2013, "Gated Tunability and Hybridization of Localized Plasmons in Nanostructured Graphene," ACS Nano **7**, 2388.

[48] Farjam, M., and H. Rafii-Tabar, 2009, "Comment on "Band structure engineering of graphene by strain: First-principles calculations"," Phys. Rev. B **80**, 167401.

[49] Faugeras, C., M. Amado, P. Kossacki, M. Orlita, M. Kühne, A. A. L. Nicolet, Yu. I. Latyshev, and M. Potemski, 2011, "Magneto-Raman Scattering of Graphene on Graphite: Electronic and Phonon Excitations," Phys. Rev. Lett. **107**, 036807.

[50] Kumar, S., 2012, "On the Thermoelectrically Pumped LED," Stanford University, http://large.stanford.edu/courses/2012/ph250/kumar2/

[51] Faugeras, C., M. Amado, P. Kossacki, M. Orlita, M. Sprinkle, C. Berger, W. A. de Heer, and M. Potemski, 2009, "Tuning the Electron-Phonon Coupling in Multilayer Graphene with Magnetic Fields," Phys. Rev. Lett. **103**, 186803.

[52] Fei, Z., *et al.*, 2012, "Gate-tuning of graphene plasmons revealed by infrared nano-imaging," Nature (London) **487**, 82.

[53] Fei, Z., *et al.*, 2013, "Electronic and plasmonic phenomena at grain boundaries in chemical vapor deposited graphene," Nat. Nanotechnol. **8**, 821.

[54] Fei, Zhe, Yi Shi, Lin Pu, Feng Gao, Yu Liu, L. Sheng, Baigeng Wang, Rong Zhang, and Youdou Zheng, 2008, "High-energy optical conductivity of graphene determined by reflection contrast spectroscopy," Phys. Rev. B **78**, 201402.

[55] Fei, Zhe, *et al.*, 2011, "Infrared Nanoscopy of Dirac Plasmons at the Graphene-SiO2 Interface," Nano Lett. **11**, 4701.

[56] Hwang, E. H., S. Adam, and S. Das Sarma, 2007a, "Carrier transport in 2D graphene layers," Phys. Rev. Lett. **98**, 186806.

[57] Hwang, E. H., S. Adam, and S. Das Sarma, 2007b, "Transport in chemically doped graphene in the presence of adsorbed molecules," Phys. Rev. B **76**, 195421.

[58] Kumar, S., Esfandyarpour, R., Davis, R., Nishi, Y., 2014, "Surface charge sensing by altering the phase transition in VO2," Journal of Applied Physics 116, 074511. http://scitation.aip.org/content/aip/journal/jap/116/7/10.1063/1.4893577

[59] Hwang, E. H., and S. Das Sarma, 2007, "Dielectric function, screening and plasmons in 2D graphene," Phys. Rev. B **75**, 205418.

[60] Hwang, E. H., and S. Das Sarma, 2008a, "Acoustic phonon scattering limited carrier mobility in two-dimensional extrinsic graphene," Phys. Rev. B **77**, 115449.

[61] Hwang, E. H., and S. Das Sarma, 2008b, "Limit to two-dimensional mobility in modulation-doped GaAs quantum structures: How to achieve a mobility of 100 million," Phys. Rev. B **77**, 235437.

[62] Hwang, E. H., and S. Das Sarma, 2008e, "Single-particle relaxation time versus transport scattering time in a two-dimensional graphene layer," Phys. Rev. B **77**, 195412.

[63] Hwang, E. H., and S. Das Sarma, 2009a, "Plasmon modes of spatially separated double-layer graphene," Phys. Rev. B **80**, 205405

[64] Kumar, S., 2012, "Types of atomic/nuclear batteries," Stanford University, http://large.stanford.edu/courses/2012/ph241/kumar1/

[65] Schuster, T., M. R. Koblischka, N. Moser, and H. Kronmüller, 1991, "Observation of nucleation and annihilation of fluxlines with opposite sign in high-Tcsuperconductors," Physica C **179**, 269

[66] Schuster, T., H. Kuhn, E.-H. Brandt, 1995, "Observation of neutral lines during flux creep in thin high-Tc superconductors," Phys. Rev. B **51**, 697

[67] Schuster, T., H. Kuhn, M. Indenbom, M. Leghissa, M. Kraus, and M. Konczykowski, 1995, "Critical-mass anisotropy due to inclined and crossed linear defects,"Phys. Rev. B **51**, 16358

[68]   Seaman, C. L., E. A. Early, M. B. Maple, W. J. Nellis, J. B. Holt, M. Kamegai, and G. S. Smith, 1989, in *Shock Compression of Condensed Matter*, edited by S. C. Schmidt, J. N. Johnson, and L. W. Davison (North-Holland, Amsterdam), p. 571

[69]   Gerber, A., and J. J. M. Franse, 1994, reply to a comment on "Self-heating versus quantum creep in bulk superconductors," Phys. Rev. Lett. **72**, 791

[70]   Suhas, K, Sripadaraja, K, 2008, "Mechanical modeling issues in optimization of dynamic behavior of RF MEMS switches," Int. J. Comp. Inform. Syst. Sci. Eng., 221-225.

[71]   Kumar, S., 2014, "Mechanisms of Resistance Switching in Various Transition Metal Oxides," Stanford University.

[72]   Gerhauser, W., G. Ries, H. W. Neumuller, W. Schmidt, O. Eibl, G. Saemann-Ischenko, and S. Klaumunzer, 1992, "Flux line pinning in Bi2Sr2CaCu2Ox crystals: Interplay of intrinsic 2D behavior and irradiation-induced columnar defects," Phys. Rev. Lett. **68**, 879

[73]   Griessen, R., 1990, "Resistive behavior of high-Tc superconductors: Influence of a distribution of activation energies," Phys. Rev. Lett. **64**, 1674

[74]   Griessen, R., 1991a, "Relaxation effects, I−V curves and irreversibility lines in high-Tc superconductors," Physica C **175**, 315

[75]   Griessen, R., 1991b, "Thermally activated flux motion near the absolute zero," Physica C **172**, 441

[76]   Griessen, R., A. Hoekstra, and R. J. Wijngaarden, 1994, Comment on "Self-heating versus quantum creep in bulk superconductors," Phys. Rev. Lett. **72**, 790

[77]   Griessen, R., J. G. Lensink, T. A. M. Schröder, and B. Dam, 1990, "Flux creep and critical currents in epitaxial high-Tc films," Cryogenics **30**, 563

[78]   Foster, Matthew S., and Igor L. Aleiner, "Graphene via large N: A renormalization group study, 2008," Phys. Rev. B **77**, 195413.

[79]   Gallagher, Patrick, Kathryn Todd, and David Goldhaber-Gordon, 2010, "Disorder-induced gap behavior in graphene nanoribbons," Phys. Rev. B **81**, 115409.

[80]   Gangadharaiah, S., A. M. Farid, and E. G. Mishchenko, 2008, "Charge Response Function and a Novel Plasmon Mode in Graphene," Phys. Rev. Lett. **100**, 166802.

[81]   Gao, Li, Jeffrey R. Guest, and Nathan P. Guisinger, 2010, "Epitaxial Graphene on Cu(111)," Nano Lett. **10**, 3512.

[82]   Suhas, K, 2010, "Materials and processes in 3D structuration of low temperature cofired ceramics for meso-scale devices," Industrial Ceramics, **30(1)**.

[83]   Grimes, C. C., and G. Adams, 1979, "Evidence for a Liquid-to-Crystal Phase Transition in a Classical, Two-Dimensional Sheet of Electrons," Phys. Rev. Lett. **42**, 795.

[84]   Groth, C. W., J. Tworzydlo, and C. W. J. Beenakker, 2008, "Electronic shot noise in fractal conductors," Phys. Rev. Lett. **100**, 176804.

[85]   Guinea, F., 2008, "Models of electron transport in single layer graphene," J. Low Temp. Phys. **153**, 359.

[86]   Gusynin, V. P., V. A. Miransky, S. G. Sharapov, and I. A. Shovkovy, 2006, "Excitonic gap, phase transition, and quantum Hall effect in graphene," Phys. Rev. B **74**, 195429.

[87]   Gusynin, V. P., and S. G. Sharapov, 2005, "Unconventional Integer Quantum Hall Effect in Graphene," Phys. Rev. Lett. **95**, 146801.

[88]   Kumar, S., 2015, "Atomic Batteries: Energy from radioactivity," arXiv:1511.07427, http://arxiv.org/abs/1511.07427

[89]   Gusynin, V. P., S. G. Sharapov, and J. P. Carbotte, 2009, "On the universal ac optical background in graphene," New J. Phys. **11**, 095013.

[90]   Aslani, M., Garner, C. M., Kumar, S., Nordlund, D., Pianetta, P., Nishi, Y., 2015, "Characterization of electronic structure of periodically strained graphene," Appl. Phys. Lett. **107 (18),** 183507.

[91]   Haldane, F. D. M., 1988, "Model for a Quantum Hall Effect without Landau Levels: Condensed-Matter Realization of the "Parity Anomaly","" Phys. Rev. Lett. **61**, 2015.

[92]   Halperin, B. I., 1982, "Quantized hall conductance, current-carrying edge states, and the existence of extended states in a two-dimensional disordered potential,"Phys. Rev. B **25**, 2185.

[93]   Han, M. Y., J. C. Brant, and P. Kim, 2010, "Electron Transport in Disordered Graphene Nanoribbons," Phys. Rev. Lett. **104**, 056801.

[94]   Han, M. Y., B. Ozyilmaz, Y. Zhang, and P. Kim, 2007, "Energy Band-Gap Engineering of Graphene Nanoribbons," Phys. Rev. Lett. **98**, 206805.

[95]   Han, W., W. H. Wang, K. Pi, K. M. McCreary, W. Bao, Y. Li, F. Miao, C. N. Lau, and R. K. Kawakami, 2009, "Electron-Hole Asymmetry of Spin Injection and Transport in Single-Layer Graphene," Phys. Rev. Lett. **102**, 137205.

[96]   Behroozi, P. S., C. Conroy, and R. H. Wechsler, 2010, "A Comprehensive Analysis of Uncertainties Affecting the Stellar Mass-Halo Mass Relation for 0<z<4,"Astrophys. J. **717**, 379.

[97] Behroozi, P. S., R. H. Wechsler, and C. Conroy, 2013, "The Average Star Formation Histories of Galaxies in Dark Matter Halos from z=0–8," Astrophys. J. **770**, 57.

[98] Bell, E. F., and R. S. de Jong, 2000, "The stellar populations of spiral galaxies," Mon. Not. R. Astron. Soc. **312**, 497.

[99] Bell, E. F., and R. S. de Jong, 2001, "Stellar Mass-to-Light Ratios and the Tully-Fisher Relation," Astrophys. J. **550**, 212.

[100] Kumar, S., 2015, "Fundamental limits to Morre's law," arXiv:1511.05956, http://arxiv.org/abs/1511.05956

[101] Gerke, A, Kumar S., Provine, J., Saraswat K., "Characterization of metal-nitride films deposited by the Savannah ALD system," Stanford University.

[102] Bell, E. F., D. H. McIntosh, N. Katz, and M. D. Weinberg, 2003, "The Optical and Near-Infrared Properties of Galaxies. I. Luminosity and Stellar Mass Functions," Astrophys. J. Suppl. Ser. **149**, 289.

[103] Bendinelli, O., 1991, "Abel integral equation inversion and deconvolution by multi-Gaussian approximation," Astrophys. J. **366**, 599.

[104] Bershady, M. A., T. P. K. Martinsson, M. A. W. Verheijen, K. B. Westfall, D. R. Andersen, and R. A. Swaters, 2011, "Galaxy Disks are Submaximal," Astrophys. J.**739**, L47.

[105] Bershady, M. A., M. A. W. Verheijen, R. A. Swaters, D. R. Andersen, K. B. Westfall, and T. Martinsson, 2010, "The DiskMass Survey. I. Overview," Astrophys. J. **716**, 198.

[106] Bershady, M. A., M. A. W. Verheijen, K. B. Westfall, D. R. Andersen, R. A. Swaters, and T. Martinsson, 2010, "The DiskMass Survey. II. Error Budget," Astrophys. J.**716**, 234.

[107] Bertin, G., L. Ciotti, and M. Del Principe, 2002, "Weak homology of elliptical galaxies.," Astron. Astrophys. **386**, 149.

[108] Kumar, S., Esfandyarpour, R., Davis, R., Nishi, Y., 2014, "Charge sensing by altering the phase transition in VO2," APS March Meeting Abstracts 1, 1135.

[109] Bertin, G., R. P. Saglia, and M. Stiavelli, 1992, "Elliptical galaxies with dark matter. I - Self-consistent models. II - Optimal luminous-dark matter decomposition for a sample of bright objects," Astrophys. J. **384**, 423.

[110] Bertola, F., D. Bettoni, L. Rusconi, and G. Sedmak, 1984, "Stellar versus gaseous kinematics in E and SO galaxies," Astron. J. **89**, 356.

[111] Bertone, G., D. Hooper, and J. Silk, 2005, "Particle dark matter: evidence, candidates and constraints," Phys. Rep. **405**, 279.

[112] Bezanson, R., *et al.*, 2011, "Redshift Evolution of the Galaxy Velocity Dispersion Function," Astrophys. J. **737**, L31.

[113] Hwang, E. H., B. Y.-K. Hu, and S. Das Sarma, 2007b, "Inelastic carrier lifetime in graphene," Phys. Rev. B **76**, 115434.

[114] Ilani, S., J. Martin, E. Teitelbaum, J. H. Smet, D. Mahalu, V. Umansky, and A. Yacoby, 2004, "The microscopic nature of localization in the quantum Hall effect," Nature (London) **427**, 328.

[115] Ishigami, M., J. H. Chen, W. G. Cullen, M. S. Fuhrer, and E. D. Williams, 2007, "Atomic structure of graphene on Si02," Nano Lett. **7**, 1643.

[116] Jayamurugan, G, Vasu, K. S., Rajesh, Y. B. R. D., Kumar, S., Vasumathi, V., Maiti, P. K., Sood, A. K., Jayaraman, N., 2011, "Interaction of single-walled carbon nanotubes with poly (propyl ether imine) dendrimers,", J. Chem. Phys. **134**, 104507.

[117] Isichenko, M. B., 1992, "Percolation, statistical topography, and transport in random-media," Rev. Mod. Phys. **64**, 961.

[118] Jackiw, R., 1984, "Fractional charge and zero modes for planar systems in a magnetic-field," Phys. Rev. D **29**, 2375.